\documentclass[prl,onecolumn,superscriptaddress,preprintnumbers,amsmath,amssymb]{revtex4}

\usepackage{bm}

\usepackage{graphicx}
\usepackage{amssymb,amsmath}
\usepackage{dcolumn}
\usepackage{bm}
\usepackage{color}
\usepackage{enumerate}

\begin{document}

\title{Control of electronic transport in graphene by electromagnetic dressing}

\author{K. Kristinsson}
\affiliation{Division of Physics and Applied Physics, Nanyang
Technological University 637371, Singapore}

\author{O. V. Kibis}\email{Oleg.Kibis@nstu.ru}
\affiliation{Department of Applied and Theoretical Physics,
Novosibirsk State Technical University, Karl Marx Avenue 20,
Novosibirsk 630073, Russia} \affiliation{Division of Physics and
Applied Physics, Nanyang Technological University 637371,
Singapore}

\author{S. Morina}
\affiliation{Division of Physics and Applied Physics, Nanyang
Technological University 637371, Singapore} \affiliation{Science
Institute, University of Iceland, Dunhagi-3, IS-107, Reykjavik,
Iceland}

\author{I. A. Shelykh}
\affiliation{Division of Physics and Applied Physics, Nanyang
Technological University 637371, Singapore} \affiliation{Science
Institute, University of Iceland, Dunhagi-3, IS-107, Reykjavik,
Iceland} \affiliation{ITMO University, St. Petersburg 197101,
Russia}

\begin{abstract}
We demonstrated theoretically that the renormalization of the
electron energy spectrum near the Dirac point of graphene by a
strong high-frequency electromagnetic field (dressing field)
drastically depends on polarization of the field. Namely, linear
polarization results in an anisotropic gapless energy spectrum,
whereas circular polarization leads to an isotropic gapped one. As
a consequence, the stationary (dc) electronic transport in
graphene strongly depends on parameters of the dressing field: A
circularly polarized field monotonically decreases the isotropic
conductivity of graphene, whereas a linearly polarized one results
in both giant anisotropy of conductivity (which can reach
thousands of percents) and the oscillating behavior of the
conductivity as a function of the field intensity. Since the
predicted phenomena can be observed in a graphene layer irradiated
by a monochromatic electromagnetic wave, the elaborated theory
opens a substantially new way to control electronic properties of
graphene with light.
\end{abstract}

\maketitle

\section{Introduction}

Since the discovery of graphene \cite{Geim_2004}, it has attracted
the persistent interest of the scientific community. Particularly,
the influence of an electromagnetic field on the electronic
properties of graphene is in the focus of attention
\cite{CastroNeto_2009,DasSarma_2011}. Usually, the electron-field
interaction is considered within the regime of weak light-matter
coupling, where the electron energy spectrum is assumed to be
unperturbed by photons. However, a lot of interesting physical
effects can be expected within the regime of strong light-matter
coupling, where the electron energy spectrum is strongly modified
by a high-frequency electromagnetic field. Following the
conventional classification, this regime is jurisdictional to
quantum optics which is an established part of modern physics
\cite{QuantumOptics1,QuantumOptics2}. Therefore, the developing of
interdisciplinary research at the border between graphene physics
and quantum optics is on the scientific agenda.

The methodology of quantum optics lies at the basis of various
exciting fields of modern physics, including quantum information
\cite{NielsenChuang}, polaritonics \cite{Kavokin}, quantum
teleportation \cite{Bennett,exp_telep}, quantum cryptography
\cite{Ekert,crypto_review}, etc. Particularly, it allows to
describe fundamental physical effects (e.g., Bose-Einstein
condensation of polaritons \cite{BEcond} and optical bistability
\cite{bistability}) and creates a basis of modern technological
applications (e.g., optical logic circuits \cite{switches}, novel
sources of terahertz emission \cite{THzLaser}, and novel types of
lasers \cite{Christopoulos,Schneider}). Within the quantum optics
approach, the system ``electron + strong electromagnetic field''
should be considered as a whole. Such a bound electron-field
system, which was called ``electron dressed by field'' (dressed
electron), became a commonly used model in modern physics
\cite{QuantumOptics1,QuantumOptics2}. The field-induced
modification of the energy spectrum and wave functions of dressed
electrons was discovered many years ago and has been studied in
detail in various atomic systems
\cite{AtomStark1,AtomStark2,AtomStark3,Kanya,Bhatia,Flegel} and
condensed matter
\cite{Goreslavskii,Mysyrowicz,Vu,Dynes,PedersenButtiker,MoskaletsButtiker,PlaterAguado,KibisStark,Kibis_2014,KibisMorina}.
In graphene-related research, the attention has been paid to the
field-induced modification of energy spectrum of dressed electrons
\cite{Naumis_2009,Oka_2009,Kibis_2010,Kibis_2011,superlattices,Calvo_2011,Calvo_2013,Syzranov_2013},
optical response of dressed electrons \cite{Zhou_2011}, transport
of dressed electrons in graphene-based p-n junctions
\cite{Fistul_2009} and electronic transport through dressed edge
states in graphene
\cite{Gu_2011,Iurov_2013,FoaTorres_2014,FoaTorres_2014_1,Dehghani_2015}.
As to stationary (dc) transport properties of a spatially
homogeneous graphene layer dressed by light, they still await
detailed analysis. The present Report is aimed to fill partially
this gap at the border between graphene physics and quantum
optics.

\section{Model}

\begin{figure}[ht]
\centering
\includegraphics[width=0.44\textwidth]{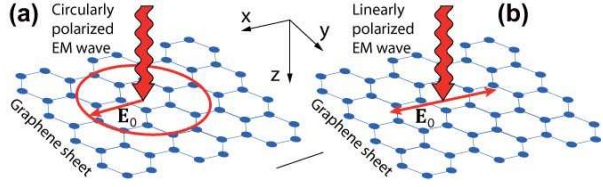}
\caption{Sketch of the electron-field system under consideration:
The graphene sheet dressed by (a) circularly polarized
electromagnetic wave with the amplitude $E_0$ and (b) linearly
polarized one.} \label{fig:system}
\end{figure}

For definiteness, we will restrict our consideration to the case
of electron states near the Dirac point of a single graphene sheet
subjected to an electromagnetic wave propagating perpendicularly
to the graphene plane. Let the graphene sheet lie in the plane
$(x,y)$ at $z=0$, and the wave propagate along the $z$ axis [see
Fig.~(\ref{fig:system})]. Then electronic properties of the
graphene are described by the Hamiltonian
\cite{CastroNeto_2009,DasSarma_2011}
\begin{equation}\label{H0}
\hat{\cal{H}}=v\bm{\sigma}\cdot(\hbar\mathbf{k}-e\mathbf{A}),
\end{equation}
where $\bm{\sigma}=(\sigma_x,\sigma_y)$ is the Pauli matrix
vector, $\mathbf{k}=(k_x,k_y)$ is the electron wave vector in the
graphene plane, $v$ is the electron velocity in graphene near the
Dirac point, $e$ is the electron charge, and
$\mathbf{A}=(A_x,A_y)$ is the vector potential of the
electromagnetic wave in the graphene plane. In what follows, we
will be to assume that the wave frequency, $\omega$, lies far from
the resonant frequencies of graphene, $2vk$. Solving the
non-stationary Schr\"odinger equation with the Hamiltonian
(\ref{H0}),
$$i\hbar\frac{\partial\psi_{\mathbf{k}}}{\partial t}=\hat{\cal{H}}\psi_{\mathbf{k}},$$
we can obtain both the energy spectrum of electrons dressed by the
electromagnetic field, $\varepsilon_{\mathbf{k}}$, and their wave
functions $\psi_{\mathbf{k}}$ (see technical details of the
solving within the Supplementary Information attached to the
Report).

For the case of the circularly polarized electromagnetic field
with the vector potential
$$\mathbf{A}=\left(\frac{E_0}{\omega}\cos\omega t,\frac{E_0}{\omega}\sin\omega t\right),$$
we arrive at the energy spectrum of the dressed electrons,
\begin{equation}\label{epsk_c}
\varepsilon_{\mathbf{k}}=\pm\sqrt{(\varepsilon_g/2)^2+(\hbar
vk)^2},
\end{equation}
where signs ``$+$'' and ``$-$'' correspond to the conduction band
and valence band of graphene, respectively,
\begin{equation}\label{Eg}
\varepsilon_g=\sqrt{\left({\hbar\omega}\right)^2+\left({2veE_0}/{\omega}\right)^2}-\hbar\omega
\end{equation}
is the field-induced band gap in graphene, $E_0$ is the amplitude
of electric field of the electromagnetic wave, and the field
frequency $\omega$ is assumed to satisfy the condition of
$\omega\gg\sqrt{2v eE_0/\hbar}$. Corresponding wave functions of
electrons dressed by the circularly-polarized field read as
\begin{equation}\label{psik_c}
\psi_{\mathbf{k}}=\varphi_{\mathbf{k}}(\mathbf{r})e^{-i\varepsilon_{\mathbf{k}}t/\hbar}
\left[\sqrt{\frac{|\varepsilon_\mathbf{k}|\mp\varepsilon_g/2}{2|\varepsilon_\mathbf{k}|}}e^{-i\theta/2}\right.
\Phi_1(\mathbf{r})
\pm\left.\sqrt{\frac{|\varepsilon_\mathbf{k}|\pm\varepsilon_g/2}{2|\varepsilon_\mathbf{k}|}}e^{i\theta/2}
\Phi_2(\mathbf{r})\right],
\end{equation}
where $\mathbf{r}=(x,y)$ is the electron radius-vector in the
graphene plane, $\Phi_{1,2}(\mathbf{r})$ are the known basic
functions of the graphene Hamiltonian (the periodical functions
arisen from atomic $\pi$-orbitals of the two crystal sublattices
of graphene) \cite{CastroNeto_2009},
$\varphi_{\mathbf{k}}(\mathbf{r})=e^{i\mathbf{k}\cdot\mathbf{r}}/\sqrt{S}$
is the plane electron wave, $S$ is the graphene area, and $\theta$
is the azimuth angle of electron in the space of wave vector,
$\mathbf{k}=(k\cos\theta,k\sin\theta)$.

In the case of linearly polarized electromagnetic field with the
vector potential
$$\mathbf{A}=\left(\frac{E_0}{\omega}\cos\omega t,0\right)$$ directed along the $x$ axis, the energy spectrum of the dressed electrons reads as
\begin{equation}\label{epsk_l}
\varepsilon_{\mathbf{k}}=\pm\hbar v k f(\theta),
\end{equation}
and the corresponding wave functions of dressed electrons are
\begin{eqnarray}\label{psik_l}
\psi_{\mathbf{k}}&=&\Bigg(\left[\Phi_1(\mathbf{r})\pm\Phi_2(\mathbf{r})\right]
\times e^{\pm i({veE_0}/{\hbar\omega^2})\sin\omega
t}-i\frac{\sin\theta}{\cos\theta+
f(\theta)}J_0\left(\frac{2veE_0}{\hbar\omega^2}\right)
\times\left[\Phi_1(\mathbf{r})\mp\Phi_2(\mathbf{r})\right]e^{\mp
i({veE_0}/{\hbar\omega^2})\sin\omega t}\Bigg)\nonumber\\
&\times&\varphi_{\mathbf{k}}(\mathbf{r})e^{-i\varepsilon_{\mathbf{k}}t/\hbar}\sqrt{\frac{\cos\theta+
f(\theta)}{4f(\theta)}},
\end{eqnarray}
where
\begin{equation}\label{f}
f(\theta)=\sqrt{\cos^2\theta+J_0^2\left(\frac{2veE_0}{\hbar\omega^2}\right)\sin^2\theta},
\end{equation}
$J_0(z)$ is the Bessel function of the first kind, and the field
frequency $\omega$ is assumed to satisfy the condition of
$\omega\gg vk$.

\begin{figure}[ht]
\centering
\includegraphics[width=0.44\textwidth]{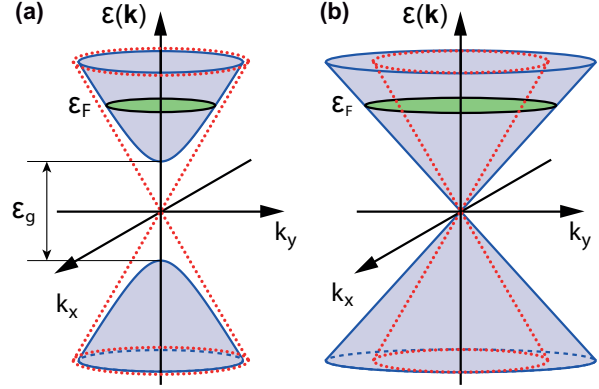}
\caption{The energy spectrum of dressed electrons in graphene for
the dressing field with different polarizations: (a) circularly
polarized dressing field; (b) dressing field polarized along the
$x$ axis. The energy spectrum of electrons in absence of the
dressing field is plotted by the dotted lines and $\varepsilon_F$
is the Fermi energy.} \label{fig:energy}
\end{figure}
The energy spectra of dressed electrons, (\ref{epsk_c}) and
(\ref{epsk_l}), are pictured schematically in
Fig.~\ref{fig:energy}. As to a consistent derivation of
Eqs.~(\ref{epsk_c})--(\ref{f}), it can be found within the
Supplementary Information attached to the Report. In order to
verify the derived expressions, it should be stressed that the
energy spectrum of electrons dressed by a classical circularly
polarized field, which is given by Eq.~(\ref{epsk_c}), exactly
coincides with the energy spectrum of electrons dressed by a
quantized field in the limit of large photon occupation numbers
\cite{Kibis_2010}. This can serve as a proof of physical
correctness of the presented approach elaborated for a classical
dressing field.

In order to calculate transport properties of dressed electrons,
we have to solve the scattering problem for nonstationary electron
states (\ref{psik_c}) and (\ref{psik_l}). Following the scattering
theory for dressed conduction electrons \cite{Kibis_2014}, the
problem comes to substituting the wave functions of dressed
electrons (\ref{psik_c}) and (\ref{psik_l}) into the conventional
expression for the Born scattering probability \cite{Landau_3}.
Assuming a total scattering potential in a graphene sheet,
$U(\mathbf{r})$, to be smooth within an elementary crystal cell of
graphene,  we can write its matrix elements as
$$
\langle\Phi_{i}(\mathbf{r})\varphi_{\mathbf{k}^\prime}(\mathbf{r})\left|U(\mathbf{r})\right|
\Phi_{j}(\mathbf{r})\varphi_{\mathbf{k}}(\mathbf{r})\rangle\approx
U_{\mathbf{k}^\prime\mathbf{k}}\delta_{ij},
$$
where
$U_{\mathbf{k}^\prime\mathbf{k}}=\langle\varphi_{\mathbf{k}^\prime}(\mathbf{r})\left|U(\mathbf{r})\right|
\varphi_{\mathbf{k}}(\mathbf{r})\rangle$, and $\delta_{ij}$ is the
Kronecker delta. As a result, the Born scattering probability for
dressed electronic states in graphene takes the form
\begin{equation}\label{W1}
w_{\mathbf{k}^\prime\mathbf{k}}=\frac{2\pi}{\hbar}
|\chi_{\mathbf{k}^\prime\mathbf{k}}|^2\left|U_{\mathbf{k}^\prime\mathbf{k}}\right|^2
\delta(\varepsilon_{{\mathbf{k}}^\prime}-\varepsilon_{\mathbf{k}}),
\end{equation}
where
\begin{equation}\label{chi_c}
\chi_{\mathbf{k}^\prime\mathbf{k}}=\sqrt{\frac{|\varepsilon_{{\bold
k}'}|-\varepsilon_g/2}{2|\varepsilon_{{\bold
k}'}|}}\sqrt{\frac{|\varepsilon_{\bold
k}|-\varepsilon_g/2}{2|\varepsilon_{\bold k}|}}
e^{i(\theta'-\theta)/2}+\sqrt{\frac{|\varepsilon_{{\bold k}'}|+
\varepsilon_g/2}{2|\varepsilon_{{\bold
k}'}|}}\sqrt{\frac{|\varepsilon_{\bold k}|+
\varepsilon_g/2}{2|\varepsilon_{\bold k}|}}
e^{-i(\theta'-\theta)/2}
\end{equation}
for the case of circularly polarized dressing field, and
\begin{equation}\label{chi_l}
\chi_{\mathbf{k}^\prime\mathbf{k}}=\sqrt{\frac{\cos(\theta')+f(\theta')}{2f(\theta')}}\sqrt{\frac{\cos(\theta)+f(\theta)}{2f(\theta)}}
\Bigg[1+\frac{\sin(\theta')}{\cos(\theta')+f(\theta')}
\frac{\sin(\theta)}{\cos(\theta)+f(\theta)}
J_0^2\left(\frac{2veE_0}{\hbar\omega^2}\right)\Bigg]
\end{equation}
for the case of linearly polarized dressing field.

In what follows, we will assume that the wave frequency, $\omega$,
meets the condition
\begin{equation}\label{omega}
\omega\tau_0\gg1,
\end{equation}
where $\tau_0$ is the electron relaxation time in an unirradiated
graphene, which should be considered as a phenomenological
parameter taken from experiments. It is well-known that the
intraband (collisional) absorption of wave energy by conduction
electrons is negligibly small under condition (\ref{omega}) (see,
e.g., Refs.~[\onlinecite{SolidState1,SolidState2,Kibis_2014}]).
Thus, the considered electromagnetic wave can be treated as a
purely dressing field which can be neither absorbed nor emitted by
conduction electrons. As a consequence, the field does not heat
the electron gas and, correspondingly, the electrons are in
thermodynamic equilibrium with a thermostat. Therefore, electron
distribution under the condition (\ref{omega}) can be described by
the conventional Fermi-Dirac function, where the energy of
``bare'' electron should be replaced with the energy of dressed
electron (\ref{epsk_c}),(\ref{epsk_l}). Substituting both this
Fermi-Dirac function and the scattering probability (\ref{W1})
into the conventional kinetic Boltzmann equation, we can analyze
the stationary (dc) transport properties of dressed electrons in
graphene. Within this approach, we take into account the two key
physical factors arisen from a dressing field: (i) modification of
the electron energy spectra (\ref{epsk_c}) and (\ref{epsk_l}) by
the dressing field; (ii) renormalization of the electron
scattering probability (\ref{W1})--(\ref{chi_l}) by the dressing
field.

\section{Results and discussion}

Let us focus our attention on the dc conductivity of the dressed
electrons. Generally, the density of the conduction electrons can
be tuned by applying a bias voltage which fixes the Fermi energy,
$\varepsilon_F$, of electron gas \cite{CastroNeto_2009}. Assuming
the Fermi energy to be in the conduction band and the temperature
to be zero, let us apply a stationary (dc) electric field
$\mathbf{E}=(E_x,E_y)$ to the graphene sheet. It follows from the
conventional Boltzmann equation for conduction electrons (see,
e.g., Refs.~[\onlinecite{Anselm,CastroNeto_2009}]) that the
electric current density, $\mathbf{J}$, is given by the expression
\begin{equation}\label{j}
\mathbf J =
\frac{e^2}{\pi^2}\int\limits_\mathbf{k}\left[\mathbf{E\cdot
v(k)}\right]\tau(\mathbf k)\mathbf v(\mathbf
k)\delta(\varepsilon_{{\mathbf{k}}}-\varepsilon_F)d^2\mathbf k,
\end{equation}
where $\mathbf{v}(\mathbf k)=({1}/{\hbar})\nabla_\mathbf
k\varepsilon_\mathbf k$ is the electron velocity, and
$\tau(\mathbf k)$ is the relaxation time. In the most general case
of anisotropic electron scattering, this relaxation time is given
by the equation \cite{Sorbello}
\begin{equation}\label{tau}
\frac{1}{\tau(\mathbf k)} = \sum_{\mathbf k'}{\left[1 -
\frac{\tau(\mathbf k')\mathbf{E\cdot v(k')}}{\tau(\mathbf
k)\mathbf{E\cdot v(k)}}\right] w_\mathbf{k'k}}.
\end{equation}
Substituting the scattering probability of dressed electron
(\ref{W1}) into Eq.~(\ref{tau}), we can obtain from
Eqs.~(\ref{j})--(\ref{tau}) the conductivity of dressed graphene,
$\sigma_{ij}=J_i/E_j$.

\begin{figure}[ht]
\centering
\includegraphics[width=0.44\textwidth]{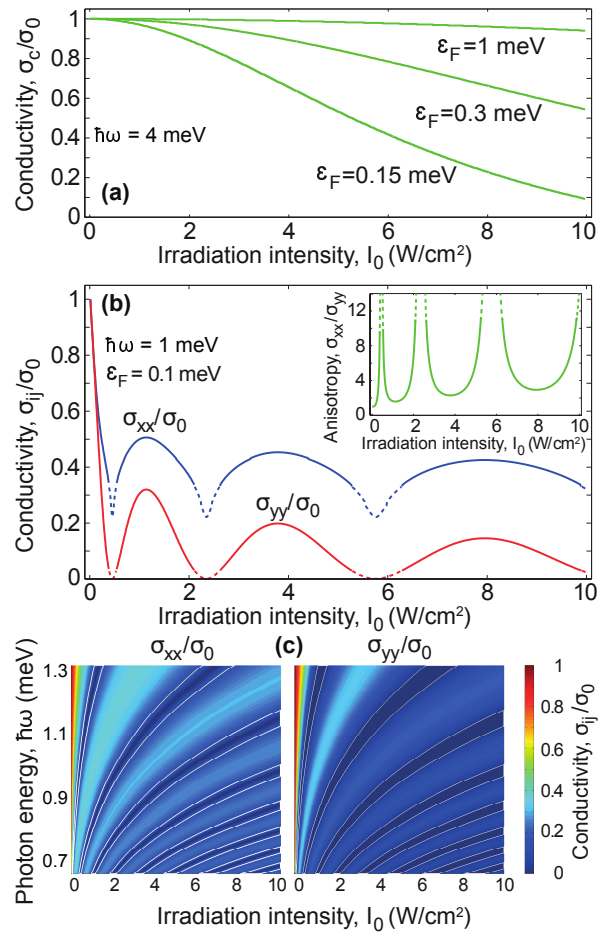}
\caption{The conductivity of dressed electrons in graphene for the
dressing field with the different polarizations: (a) circularly
polarized dressing field; (b)--(c) dressing field polarized along
the $x$ axis. Physically relevant regions of the field parameters,
where the developed theory is applicable, correspond to the solid
lines in the plot (b) and wide areas between the dashed lines in
the plot (c).} \label{fig:conductivity}
\end{figure}

To simplify calculations, let us consider the electron scattering
within the $s$-wave approximation \cite{Landau_3}, where the
matrix elements $U_{\mathbf{k}^\prime\mathbf{k}}$ do not depend on
the angle
$\theta_{\mathbf{k}^\prime\mathbf{k}}=(\widehat{\mathbf{k}^\prime,\mathbf{k}})$.
Substituting the probability (\ref{W1}) into
Eqs.~(\ref{j})--(\ref{tau}), we arrive at the isotropic
conductivity of a graphene dressed by a circularly polarized
field, $\sigma_c=\sigma_{xx}=\sigma_{yy}$, which is given by the
expression
\begin{equation}\label{tau_c}
\frac{\sigma_c}{\sigma_0}=
\frac{1-{\varepsilon_g^2}/{4\varepsilon_F^2}}{1+{3\varepsilon_g^2}/{4\varepsilon_F^2}},
\end{equation}
where $\varepsilon_F\geq\varepsilon_g$. It is seen in
Fig.~\ref{fig:conductivity}a that the conductivity (\ref{tau_c})
monotonically decreases with increasing field intensity
$I_0=\epsilon_0E_0^2c/2$. Physically, this behavior is a
consequence of decreasing Fermi velocity, $\mathbf{v}_F(\mathbf
k)=({1}/{\hbar})\nabla_\mathbf k\varepsilon_\mathbf
k|_{\varepsilon=\varepsilon_F}$, with increasing field amplitude
$E_0$ (see Fig.~\ref{fig:energy}a). For the case of a dressing
field linearly polarized along the $x$ axis, the conductivity is
plotted in Figs.~\ref{fig:conductivity}b--\ref{fig:conductivity}c.
There are the two main features of the conductivity as a function
of the dressing field intensity: Firstly, the conductivity
oscillates, and, secondly, the giant anisotropy of the
conductivity, $\sigma_{xx}/\sigma_{yy}\sim10$ appears (see the
insert in Fig.~\ref{fig:conductivity}b). The oscillating behavior
arises from the Bessel functions which take place in both the
energy spectrum (\ref{epsk_l}) and the scattering probability
(\ref{W1}). As to the conductivity anisotropy, it is caused by the
field-induced anisotropy of the energy spectrum (\ref{epsk_l}).
Namely, the linearly polarized dressing field turns the round
(isotropic) Fermi line of unperturbed graphene into the strongly
anisotropic ellipse line (see Fig.~\ref{fig:energy}b). As a
result, the Fermi velocities of dressed electrons along the $x,y$
axes are strongly different and the discussed anisotropic
conductivity appears. It should be stressed that the
aforementioned features of electronic properties are typical
exclusively for linear electron dispersion and, correspondingly,
do not take place in a dressed electron gas with parabolic
dispersion \cite{KibisMorina}. To avoid misunderstandings, it
should be noted also that the zeros of conductivity in
Fig.~\ref{fig:conductivity}b lie within physically irrelevant
areas pictured by dashed lines. Formally, these irrelevant areas
correspond to the broken condition $\omega\gg vk_F$, which is
crucial for the correctness of the energy spectrum (\ref{epsk_l})
at the Fermi energy. Thus, the zeros have no physical meaning and
should be ignored.

It is seen in Fig.~\ref{fig:conductivity} that the behavior of
conductivity is qualitatively different for the dressing field
with different polarizations. Physically, the strong polarization
dependence of electronic transport follows directly from the
strong polarization dependence of energy spectrum of dressed
electrons. Namely, the energy spectrum of electrons dressed by a
circularly polarized field (\ref{epsk_c}) is isotropic and has the
field-induced gap (\ref{Eg}) at the Dirac point. In contrast, the
energy spectrum of electrons dressed by the linearly polarized
field (\ref{epsk_l}) is gapless and has the field-induced
anisotropy arisen from the Bessel function in Eq.~(\ref{f}). These
differences in the spectra (\ref{epsk_c}) and (\ref{epsk_l}) lead
to the discussed difference of transport for electrons dressed by
circularly polarized light and linearly polarized one. It should
be noted that an electromagnetic field can open energy gaps within
conduction and valence bands at electron wave vectors
$\mathbf{k}\neq0$ (see, e.g.,
Refs.~[\onlinecite{superlattices,Calvo_2011,Calvo_2013,Syzranov_2013}]).
These gaps arise from the optical (ac) Stark effect and take place
at resonant points of the Brillouin zone, where the condition of
$\omega=2vk$ is satisfied. Certainly, the basic expressions
(\ref{epsk_c})--(\ref{f}) are not applicable near the Stark gaps.
However, these gaps lie far from the Dirac point in the case of
high-frequency dressing field. Therefore, they do not influence on
low-energy electronic transport under consideration.

\section{Conclusions}

We have shown that the transport properties of electrons in
graphene are strongly affected by a dressing field. Namely, a
circularly polarized dressing field monotonically decreases the
isotropic conductivity of graphene, whereas a linearly polarized
dressing field results in the oscillating behavior of the
conductivity and its giant anisotropy. As a result, the dc
transport properties of graphene can be effectively controlled by
a strong high-frequency electromagnetic field. From the viewpoint
of possible applications, the discussed effect can make graphene
more tunable. Particularly, the switching times for conductivity
of graphene controlled by a high-frequency field are expected to
be shorter then for the case of conventional electrostatic control
of conductivity by gate electrodes. This can create physical
prerequisites for novel graphene-based optoelectronic devices.

\section{Acknowledgements}

The work was partially supported by FP7 IRSES projects POLATER and
QOCaN, FP7 ITN project NOTEDEV, Rannis project BOFEHYSS, RFBR
project 14-02-00033, the Russian Target Federal Program (project
14.587.21.0020) and the Russian Ministry of Education and Science.

\section{Author contributions}

O.V.K. and I.A.S. formulated the physical problem under
consideration and derived analytical solutions of the problem.
K.K. and S.M. analyzed the basic expressions describing the
problem, performed numerical calculations and plotted figures.
O.V.K. and K.K. wrote the paper. All co-authors taken part in
discussions of used physical models and obtained results.

\section{Additional information}

\textbf{Competing financial interests:} The authors declare no
competing financial interests.

\section*{SUPPLEMENTARY INFORMATION:\\ Full and consistent derivation of the energy
spectrum and wave functions of dressed electrons in graphene}

Let us consider a graphene sheet which lies in the plane $(x,y)$
at $z=0$ and is subjected to an electromagnetic wave propagating
along the $z$ axis (dressing electromagnetic field). Then
electronic properties of the graphene are described by the
Hamiltonian
\begin{equation}\label{AH0}
\hat{\cal{H}}=v\bm{\sigma}\cdot(\hbar\mathbf{k}-e\mathbf{A}),
\end{equation}
where $\bm{\sigma}=(\sigma_x,\sigma_y)$ is the Pauli matrix
vector, $\mathbf{k}=(k_x,k_y)$ is the electron wave vector in the
graphene plane, $v$ is the electron velocity in graphene near the
Dirac point, $e$ is the electron charge, and
$\mathbf{A}=(A_x,A_y)$ is the vector potential of the
electromagnetic wave in the graphene plane. Solving the
Schr\"odinger equation with the Hamiltonian (\ref{AH0}), we can
find the energy spectrum of dressed electrons and their wave
functions as follows.

\subsection*{I. Circularly polarized dressing field}

For the case of circularly polarized electromagnetic wave, its
vector potential $\mathbf{A}=(A_x,A_y)$ can be written as
\begin{equation}\label{AAc}
\mathbf{A}=\left(\frac{E_0}{\omega}\cos\omega
t,\frac{E_0}{\omega}\sin\omega t\right),
\end{equation}
where $E_0$ is the electric field amplitude of the wave, and
$\omega$ is the wave frequency. Then the Hamiltonian (\ref{AH0})
is
\begin{equation}\label{AHomega}
\hat{\cal{H}}=\hat{\cal{H}}_0+\hat{\cal{H}}_k,
\end{equation}
where
\begin{equation}\label{AH00}
\hat{\cal{H}}_0=\left(-\frac{veE_0}{\omega}\right)
\begin{pmatrix}
0 & e^{-i\omega t}\\
e^{i\omega t} & 0
\end{pmatrix},
\end{equation}
and
\begin{equation}\label{Hk}
\hat{\cal{H}}_k=
\begin{pmatrix}
0 & v(\hbar k_x-i\hbar k_y)\\
v(\hbar k_x+i\hbar k_y) & 0
\end{pmatrix}.
\end{equation}
The nonstationary Schr\"odinger equation with the Hamiltonian
(\ref{AH00}),
\begin{equation}\label{Ashr1}
i\hbar\frac{\partial\psi_0}{\partial t}=\hat{\cal{H}}_0\psi_0,
\end{equation}
describes the time evolution of electron states in the Dirac point
($\mathbf{k}=0$). The exact solutions of the equation
(\ref{Ashr1}) can be sought in the form
\begin{equation}\label{Apsi1}
\psi_0=e^{-i\alpha t/\hbar}[A\Phi'_1(\mathbf{r})e^{-i\omega
t/2}+B\Phi'_2(\mathbf{r})e^{i\omega t/2}],
\end{equation}
where $\Phi'_{1,2}(\mathbf{r})$ are the basic functions of the
$2\times2$ matrix Hamiltonian (\ref{AH0}), and $\alpha$, $A$ and
$B$ are the sought parameters. Substituting the wave function
(\ref{Apsi1}) into the Schr\"odinger equation (\ref{Ashr1}), we
arrive at the system of algebraic equations
\begin{eqnarray}\label{AAB1}
A\left(\alpha+\frac{\hbar\omega}{2}\right)+B\left(\frac{veE_0}{\omega}\right)&=&0\nonumber\\
A\left(\frac{veE_0}{\omega}\right)+B\left(\alpha-\frac{\hbar\omega}{2}\right)&=&0.
\end{eqnarray}
The condition of nontrivial solution of the system (\ref{AAB1}),
$$\begin{vmatrix}
\alpha+\frac{\hbar\omega}{2} & \frac{veE_0}{\omega}\\
\frac{veE_0}{\omega} & \alpha-\frac{\hbar\omega}{2}
\end{vmatrix}=0,$$
gives the two different parameters, $\alpha=\pm\Omega/2$, where
\begin{equation}
\Omega=\sqrt{\left({\hbar\omega}\right)^2+\left(\frac{2veE_0}{\omega}\right)^2}.
\end{equation}
Therefore, there are two sets of solutions of the system
(\ref{AAB}), which correspond to these two parameters and satisfy
the normalization condition, $|A|^2+|B|^2=1$. As a result, there
are two wave functions (\ref{Apsi1}),
\begin{equation}\label{Apsi21}
\psi_{0}^{\pm}={e^{\pm i\Omega
t/2\hbar}}\left[\sqrt{\frac{\Omega\pm\hbar\omega}{2\Omega}}\Phi'_1(\mathbf{r})e^{
-i\omega t/2}\right.
\pm\left.\frac{e}{|e|}\sqrt{\frac{\Omega\mp\hbar\omega}{2\Omega}}\Phi'_2(\mathbf{r})e^{i\omega
t/2}\right],
\end{equation}
which exactly describe electron states of irradiated graphene in
the Dirac point ($\mathbf{k}=0$). Since the two wave functions
(\ref{Apsi21}) are the complete function system for any time $t$,
we can seek the solution of the Schr\"odinger equation with the
full Hamiltonian (\ref{AHomega}) as an expansion
\begin{equation}\label{Apsik1}
\psi_{\mathbf{k}}=a^+(t)\psi_{0}^{+}+a^-(t)\psi_{0}^{-}.
\end{equation}
Substituting the expansion (\ref{Apsik1}) into the Schr\"odinger
equation with the full Hamiltonian (\ref{AHomega}),
\begin{equation}\label{Ashr11}
i\hbar\frac{\partial\psi_{\mathbf{k}}}{\partial
t}=\hat{\cal{H}}\psi_{\mathbf{k}},
\end{equation}
we arrive at the system of two differential equations for the
coefficients $a^\pm(t)$,
\begin{eqnarray}\label{Aa1}
i\dot{a}^+(t)&=&v\frac{e}{|e|}\left[\frac{W_0}{\Omega}(k_x\cos\omega
t+k_y\sin\omega t)a^+(t)\right.
-\left.(k_x-ik_y)\left(\frac{\Omega+\hbar\omega}{2\Omega}\right)e^{-i(\Omega/\hbar-\omega)t}a^-(t)\right.\nonumber\\
&+&\left.(k_x+ik_y)\left(\frac{\Omega-\hbar\omega}{2\Omega}\right)e^{-i(\Omega/\hbar+\omega)t}a^-(t)\right],\nonumber\\
i\dot{a}^-(t)&=&-v\frac{e}{|e|}\left[\frac{W_0}{\Omega}(k_x\cos\omega
t+k_y\sin\omega t)a^-(t)\right.
+\left.(k_x+ik_y)\left(\frac{\Omega+\hbar\omega}{2\Omega}\right)e^{i(\Omega/\hbar-\omega)t}a^+(t)\right.\nonumber\\
&-&\left.(k_x-ik_y)\left(\frac{\Omega-\hbar\omega}{2\Omega}\right)e^{i(\Omega/\hbar+\omega)t}a^+(t)\right],
\end{eqnarray}
where $$W_0=\frac{2veE_0}{\omega}$$ is the characteristic kinetic
energy of rotational electron motion induced by the circularly
polarized field. In what follows, we will assume that the field
frequency $\omega$ is high enough to satisfy the condition
\begin{equation}\label{Acc}
W_0/\hbar\omega\ll1.
\end{equation}
Then we have
\begin{equation}\label{Aww}
\frac{W_0}{\Omega}\approx\frac{W_0}{\hbar\omega}\approx0,\,\,\frac{\Omega+\hbar\omega}{2\Omega}\approx1,\,\,\frac{\Omega-\hbar\omega}{2\Omega}\approx0,
\end{equation}
and, correspondingly, Eqs.~(\ref{Aa1}) take the form
\begin{eqnarray}\label{Aaa}
i\dot{a}^+(t)&=&-v\frac{e}{|e|}(k_x-ik_y)e^{-i(\Omega-\hbar\omega)t/\hbar}a^-(t),\nonumber\\
i\dot{a}^-(t)&=&-v\frac{e}{|e|}(k_x+ik_y)e^{i(\Omega-\hbar\omega)t/\hbar}a^+(t).
\end{eqnarray}
Let us seek solutions of Eqs.~(\ref{Aaa}) as $a_c^\pm(t)=C^\pm
e^{-i(\pm\Omega/2\mp\hbar\omega/2+\varepsilon_{\mathbf{k}})t/\hbar}$,
where $C^\pm$ and $\varepsilon_{\mathbf{k}}$ are the sought
parameters. Solving the system of equations (\ref{Aaa}), we arrive
at
\begin{equation}\label{AEE}
\varepsilon_{\mathbf{k}}=\pm\sqrt{(\varepsilon_g/2)^2+(\hbar
vk)^2},
\end{equation}
and
\begin{equation}\label{Aeg}
\varepsilon_g=\sqrt{\left({\hbar\omega}\right)^2+\left(\frac{2veE_0}{\omega}\right)^2}-\hbar\omega.
\end{equation}
Following the conventional terminology, Eq.~(\ref{AEE}) describes
the spectrum of quasienergies of electrons in graphene dressed by
a circularly polarized field, where signs ``$+$'' and ``$-$''
correspond to conduction band and valence band, respectively. The
two wave functions (\ref{Apsik1}), which correspond to the two
energy branches (\ref{AEE}), can be written as
\begin{eqnarray}\label{Apsikc}
\psi_{\mathbf{k}}&=&e^{-i\varepsilon_{\mathbf{k}}
t/\hbar}\left[\pm\sqrt{\frac{|\varepsilon_{\mathbf{k}}|\mp\varepsilon_g/2}{2|\varepsilon_{\mathbf{k}}|}}e^{-i\theta/2}\right.
\left(\frac{e}{|e|}\sqrt{\frac{\Omega+\hbar\omega}{2\Omega}}\Phi'_1(\mathbf{r})+\sqrt{\frac{\Omega-\hbar\omega}{2\Omega}}\Phi'_2(\mathbf{r})e^{
i\omega t}\right)\nonumber\\
&-&\sqrt{\frac{|\varepsilon_{\mathbf{k}}|\pm\varepsilon_g/2}{2|\varepsilon_{\mathbf{k}}|}}e^{i\theta/2}
\left(\sqrt{\frac{\Omega-\hbar\omega}{2\Omega}}\Phi'_1(\mathbf{r})e^{-i\omega
t}\right.
-\left.\left.\frac{e}{|e|}\sqrt{\frac{\Omega+\hbar\omega}{2\Omega}}\Phi'_2(\mathbf{r})\right)\right],
\end{eqnarray}
where $\theta$ is the azimuth angle of the wave vector,
$\mathbf{k}=(k\cos\theta,k\sin\theta)$. Let us write the basis
functions $\Phi'_{1,2}(\mathbf{r})$ in the conventional Bloch
form,
$\Phi'_{1,2}(\mathbf{r})=\Phi_{1,2}(\mathbf{r})\varphi_{\mathbf{k}}(\mathbf{r})$,
where $\Phi_{1,2}(\mathbf{r})$ are the periodical functions arisen
from atomic $\pi$-orbitals of the two crystal sublattices of
graphene,
$\varphi_{\mathbf{k}}(\mathbf{r})=e^{i\mathbf{k}\cdot\mathbf{r}}/\sqrt{S}$
is the plane electron wave, and $S$ is the graphene area. Keeping
in mind inequalities (\ref{Aww}), we arrive from (\ref{Apsikc}) to
the sought wave functions of dressed electrons in the final form,
\begin{equation}\label{Apsikcf}
\psi_{\mathbf{k}}=\varphi_{\mathbf{k}}(\mathbf{r})e^{-i\varepsilon_{\mathbf{k}}t/\hbar}
\left[\sqrt{\frac{|\varepsilon_{\mathbf{k}}|\mp\varepsilon_g/2}{2|\varepsilon_{\mathbf{k}}|}}e^{-i\theta/2}\right.
\Phi_1(\mathbf{r})\pm
\left.\sqrt{\frac{|\varepsilon_{\mathbf{k}}|\pm\varepsilon_g/2}{2|\varepsilon_{\mathbf{k}}|}}e^{i\theta/2}
\Phi_2(\mathbf{r})\right].
\end{equation}

\subsection*{II. Linearly polarized dressing field}

For the case of electromagnetic wave linearly polarized along the
$x$ axis, its vector potential $\mathbf{A}=(A_x,A_y)$ can be
written as
\begin{equation}\label{AAl}
\mathbf{A}=\left(\frac{E_0}{\omega}\cos\omega t,0\right).
\end{equation}
Then the Hamiltonian (\ref{AH0}) is
\begin{equation}\label{AHomega2}
\hat{\cal{H}}=\hat{\cal{H}}_0+\hat{\cal{H}}_k,
\end{equation}
where
\begin{equation}\label{AH002}
\hat{\cal{H}}_0=\begin{pmatrix}
0 & -1\\
-1 & 0
\end{pmatrix}\frac{veE_0}{\omega}\cos\omega t.
\end{equation}
and
\begin{equation}\label{AHk2}
\hat{\cal{H}}_k=
\begin{pmatrix}
0 & v(\hbar k_x-i\hbar k_y)\\
v(\hbar k_x+i\hbar k_y) & 0
\end{pmatrix}.
\end{equation}
The nonstationary Schr\"odinger equation with the Hamiltonian
(\ref{AH002}),
\begin{equation}\label{Ashr12}
i\hbar\frac{\partial\psi_0}{\partial t}=\hat{\cal{H}}_0\psi_0,
\end{equation}
describes the time evolution of electron states in the Dirac point
($\mathbf{k}=0$). The exact solutions of this Schr\"odinger
equation can be sought in the form
\begin{equation}\label{Apsil1}
\psi_0=[A\Phi'_1(\mathbf{r})+B\Phi'_2(\mathbf{r})]
\exp\left[-i\alpha\frac{veE_0}{\hbar\omega^2}\sin\omega t\right],
\end{equation}
where $\alpha$, $A$ and $B$ are the sought parameters.
Substituting the wave function (\ref{Apsil1}) into the
Schr\"odinger equation (\ref{Ashr12}) with the Hamiltonian
(\ref{AH002}), we arrive at the system of algebraic equations
\begin{eqnarray}\label{AAB}
A\alpha+B&=&0\nonumber\\
A+B\alpha&=&0.
\end{eqnarray}
The condition of nontrivial solution of the system (\ref{AAB}),
$$\begin{vmatrix}
\alpha & 1\\
1 & \alpha
\end{vmatrix}=0,$$
gives the two different parameters, $\alpha=\pm1$. Therefore,
there are two sets of solutions of the system (\ref{AAB}), which
correspond to these two parameters and satisfy the normalization
condition, $|A|^2+|B|^2=1$:
\begin{eqnarray}
A=B=\frac{1}{\sqrt{2}},\nonumber\\
A=-B=\frac{1}{\sqrt{2}}.
\end{eqnarray}
As a result, there are two wave functions (\ref{Apsil1}),
\begin{equation}\label{Apsi2}
\psi_0^\pm=\frac{1}{\sqrt{2}}
\left[\Phi'_1(\mathbf{r})\pm\Phi'_2(\mathbf{r})\right]\exp\left[\pm
i\frac{veE_0}{\hbar\omega^2}\sin\omega t\right],
\end{equation}
which exactly describe electron states of irradiated graphene in
the Dirac point ($\mathbf{k}=0$). Since the two wave functions
(\ref{Apsi2}) are the complete function system for any time $t$,
we can seek the solution of the nonstationary Schr\"odinger
equation with the full Hamiltonian (\ref{AHomega2}) as an
expansion
\begin{equation}\label{Apsik}
\psi_{\mathbf{k}}=a^+(t)\psi^+_0+a^-(t)\psi^-_0.
\end{equation}
Substituting the expansion (\ref{Apsik}) into this Schr\"odinger
equation with the full Hamiltonian (\ref{AHomega2}),
\begin{equation}\label{Ashr112}
i\hbar\frac{\partial\psi_{\mathbf{k}}}{\partial
t}=\hat{\cal{H}}\psi_{\mathbf{k}},
\end{equation}
we arrive at the system of two differential equations for the
coefficients $a^\pm(t)$,
\begin{eqnarray}\label{Aa}
i\dot{a}^+(t)&=&vk_xa^+(t)+ivk_ya^-(t)\exp\left[-
i\frac{2veE_0}{\hbar\omega^2}\sin\omega t\right],\nonumber\\
i\dot{a}^-(t)&=&-ivk_ya^+(t)\exp\left[
i\frac{2veE_0}{\hbar\omega^2}\sin\omega t\right]-vk_xa^-(t).\nonumber\\
\end{eqnarray}
It follows from the Floquet's theorem that the functions
$a^\pm(t)$ can be written as
\begin{equation}\label{Afl}
a^\pm(t)=e^{-i\varepsilon_{\mathbf{k}}
t/\hbar}\widetilde{a}^\pm(t),
\end{equation}
where $\varepsilon_{\mathbf{k}}$ is the electron quasienergy in
the irradiated graphene (the energy spectrum of dressed
electrons), and $\widetilde{a}^\pm(t)$ are the periodical
functions with the period $T=2\pi/\omega$. Let us apply the
Jacobi-Anger expansion,
$$e^{iz\sin\gamma}=\sum_{n=-\infty}^{\infty}J_n(z)e^{in\gamma},$$
to the exponents in the right side of Eqs.~(\ref{Aa}) and expand
the periodical functions $\widetilde{a}^\pm(t)$ into the Fourier
series
$$\widetilde{a}^\pm(t)=\sum_{n=-\infty}^{\infty}c^\pm_n e^{in\omega t}.$$
Then the differential equations (\ref{Aa}) can be transformed to
the algebraic equations
\begin{eqnarray}\label{Aa3}
\left(vk_x-\frac{\varepsilon_{\mathbf{k}}}{\hbar}-n\omega\right)c^+_n&+&ivk_y\sum_{m=-\infty}^{\infty}c^-_{n-m}J_m\left(-\frac{2veE_0}{\hbar\omega^2}\right)=0,\nonumber\\
\left(vk_x+\frac{\varepsilon_{\mathbf{k}}}{\hbar}+n\omega\right)c^-_n&+&ivk_y\sum_{m=-\infty}^{\infty}c^+_{n-m}J_m\left(\frac{2veE_0}{\hbar\omega^2}\right)=0,
\end{eqnarray}
where $J_m(z)$ is the Bessel function of the first kind.
Equations~(\ref{Aa3}) can be easily solved in the case of
high-frequency field satisfying the conditions
\begin{equation}\label{Ac2}
\hbar\omega\gg\varepsilon_{\mathbf{k}},\,\,\omega\gg vk.
\end{equation}
In this simplest case, Eqs.~(\ref{Aa3}) for $n\neq0$ can be
reduced to the equation
\begin{equation}\label{Aa5}
c^\pm_n\approx
i\frac{vk_y}{n\omega}\sum_{m=-\infty}^{\infty}c^\mp_{n-m}J_m\left(\mp\frac{2veE_0}{\hbar\omega^2}\right).
\end{equation}
Keeping in mind that $|c^\pm_n|\leq1$ and $|J_n(z)|\leq1$,
Eq.~(\ref{Aa5}) leads to the evident solution
$c^\pm_{n\neq0}\approx0$. After substitution of this solution into
Eqs.~(\ref{Aa3}), the expressions (\ref{Aa3}) are reduced to the
two equations
\begin{eqnarray}\label{Aa4}
\left(vk_x-\frac{\varepsilon_{\mathbf{k}}}{\hbar}\right)c^+_0+ivk_yc^-_0J_0\left(\frac{2veE_0}{\hbar\omega^2}\right)&=&0,\nonumber\\
\left(vk_x+\frac{\varepsilon_{\mathbf{k}}}{\hbar}\right)c^-_0+ivk_yc^+_0J_0\left(\frac{2veE_0}{\hbar\omega^2}\right)&=&0.\nonumber\\
\end{eqnarray}
Solving the system of the two algebraic equations (\ref{Aa4})
accurately, we can easily obtain both the coefficients $c^\pm_0$
and the energy spectrum of dressed electrons,
\begin{equation}\label{AEn}
\varepsilon_{\mathbf{k}}=\pm\hbar vk f(\theta),
\end{equation}
where
$$f(\theta)=\sqrt{\cos^2\theta+J_0^2\left(\frac{2veE_0}{\hbar\omega^2}\right)\sin^2\theta}.$$
Correspondingly, the sought wave functions of dressed electrons
(\ref{Apsik}) take the final form
\begin{eqnarray}\label{Apsiklf}
\psi_{\mathbf{k}}&=&\varphi_{\mathbf{k}}(\mathbf{r})e^{-i\varepsilon_{\mathbf{k}}t/\hbar}\sqrt{\frac{\cos\theta+
f(\theta)}{4f(\theta)}}\Bigg(\left[\Phi_1(\mathbf{r})\pm\Phi_2(\mathbf{r})\right]
e^{\pm i({veE_0}/{\hbar\omega^2})\sin\omega
t}\nonumber\\
&-&i\frac{\sin\theta}{\cos\theta+
f(\theta)}J_0\left(\frac{2veE_0}{\hbar\omega^2}\right)
\left[\Phi_1(\mathbf{r})\mp\Phi_2(\mathbf{r})\right]e^{\mp
i({veE_0}/{\hbar\omega^2})\sin\omega t}\Bigg).
\end{eqnarray}
Taking into account Eq.~(\ref{AEn}), the two conditions
(\ref{Ac2}) can be reduced to the solely condition,
\begin{equation}\label{Ac3}
\omega\gg vk,
\end{equation}
which describes borders of applicability of Eqs.~(\ref{AEn}) and
(\ref{Apsiklf}).

\end{document}